\documentclass[pra,twocolumn,aps,amssymb,amsmath,superscriptaddress,showpacs]{revtex4-1}
\usepackage[latin1]{inputenc}
\usepackage{mathrsfs}
\usepackage{dsfont}
\usepackage{amsmath}
\usepackage{amssymb}
\usepackage{amsthm}
\usepackage{amsfonts}
\usepackage{amstext}
\usepackage{amsopn}
\usepackage{amsxtra}
\usepackage{mathrsfs}
\usepackage{dsfont}
\usepackage{color}
\usepackage{graphicx}
\newtheorem{theorem}{Theorem}
\newtheorem{lemma}{Lemma}
\newcommand{\dps}{\displaystyle}
\newcommand{\ii}{\infty}
\newcommand\R{{\ensuremath {\mathbb R} }}

\newcommand\N{{\ensuremath {\mathbb N} }}
\newcommand\Z{{\ensuremath {\mathbb Z} }}

\newcommand\1{{\ensuremath {\mathds 1} }}

\newcommand\nn{\nonumber}

\renewcommand\phi{\varphi}

\newcommand{\bP}{\mathbb{P}}

\newcommand{\wto}{\rightharpoonup}

\newcommand{\cE}{\mathcal{E}}

\newcommand{\cL}{\mathcal{L}}

\newcommand{\bx}{\mathbf{x}}
\newcommand{\by}{\mathbf{y}}
\newcommand{\bz}{\mathbf{z}}
\newcommand{\ba}{\mathbf{a}}

\newcommand{\br}{\mathbf{r}}
\newcommand{\bp}{\mathbf{p}}

\newcommand{\bk}{\mathbf{k}}

\newcommand{\dr}{{\rm d}r}

\renewcommand{\epsilon}{\varepsilon}

\newcommand{\norm}[1]{ \left| \! \left| #1 \right| \! \right| }

\renewcommand{\geq}{\geqslant}
\renewcommand{\leq}{\leqslant}

\renewcommand{\tilde}{\widetilde}

\newcommand{\dx}{{\rm d}\bx}
\renewcommand{\dr}{{\rm d}\br}
\newcommand{\dy}{{\rm d}\by}
\newcommand{\dz}{{\rm d}\bz}

\newcommand{\dpp}{{\rm d}\bp}

\newcommand{\da}{{\rm d}\ba}

\date{\today}

\begin{document}

\title{Floating Wigner crystal with no boundary charge fluctuations}

\author{Mathieu Lewin}
\affiliation{CNRS \& CEREMADE, Paris-Dauphine University, PSL University, 75016 Paris, France}

\author{Elliott H. Lieb}
\affiliation{Departments of Physics and Mathematics, Princeton University, Jadwin Hall, Washington Road, Princeton, NJ 08544, USA}

\author{Robert Seiringer}
\affiliation{IST Austria (Institute of Science and Technology Austria), Am Campus 1, 3400 Klosterneuburg, Austria}

\begin{abstract}
We modify the ``floating crystal'' trial state for the classical Homogeneous Electron Gas (also known as Jellium), in order to suppress the boundary charge fluctuations that are known to lead to a macroscopic increase of the energy. The  argument is to melt a thin layer of the crystal close to the boundary and consequently replace it by an incompressible fluid. With the aid of this trial state we show that three different definitions of the ground state energy of Jellium coincide. In the first point of view the electrons are placed in a neutralizing uniform background. In the second definition there is no background but the electrons are submitted to the constraint that their density is constant, as is appropriate in Density Functional Theory. Finally, in the third system each electron interacts with a periodic image of itself, that is, periodic boundary conditions are imposed on the interaction potential.
\end{abstract}

\maketitle

\section{Introduction}

The \emph{Homogeneous Electron Gas}, also called \emph{Jellium}, is a fundamental system in quantum physics and chemistry~\cite{ParYan-94,GiuVig-05}. In this paper we 
introduce a modified ``floating crystal'' trial state and use it to prove that three possible definitions of the Jellium ground state energy coincide in the thermodynamic limit. In particular, we resolve a conundrum originating in~\cite{NavJamFei-80,ChoFavGru-80,BorBorShaZuc-88,BorBorSha-89,BorBorStr-14} and raised again in~\cite{LewLie-15}, where it was observed that the usual floating crystal trial state fails for Coulomb interactions.

In its original formulation, due to Wigner~\cite{Wigner-34}, Jellium is defined as an infinite gas of electrons placed in a positively-charged uniform background. The thermodynamic limit of this system has been rigorously established in~\cite{LieNar-75}. This model provides a good description of the deep interior of white dwarfs~\cite{Salpeter-61,BauHan-80} (where the point charges are the fully ionized atoms evolving in a uniform background of negatively charged electrons). It has also been shown to be of high relevance for valence electrons in alkaline metals, for instance in solid sodium~\cite{Huotari-etal-10}. 

A similar system appears in the Local Density Approximation of Density Functional Theory (DFT)~\cite{ParYan-94}, where it plays a central role for deriving functionals~\cite{Perdew-91,PerWan-92,Becke-93,PerBurErn-96}. In DFT the density is fixed and there is no background. The natural system arising in this situation is an infinite gas of electrons submitted to the \emph{constraint} that its density is constant over the whole space. This model was called the \emph{Uniform Electron Gas} (UEG) in~\cite{LewLie-15,LewLieSei-18,LewLieSei-19} to avoid any possible confusion with Jellium. 

At low density, the electrons in Jellium are believed to form a BCC Wigner crystal~\cite{Wigner-34,Wigner-38}, hence their density is not at all constant. It has nevertheless been assumed by many authors that the two definitions should coincide. The reason for this belief is that the Wigner crystal has no preferred position and orientation, hence one may consider the mixed state obtained by uniformly averaging over the position of the lattice. This state is sometimes called the \emph{floating crystal}~\cite{BisLuh-82,MikZie-02,DruRadTraTowNee-04} and it has a constant density. 

With long range potentials such as Coulomb, one should however be very careful, since boundary effects can easily play a decisive role. It was proved in~\cite{LewLie-15} that, for the classical gas, computing the energy of the floating crystal in a thermodynamic limit leads to a much higher energy than the Jellium energy of the BCC crystal, with a shift of the order of the volume of the sample due to charge fluctuations close to the boundary. This is very specific to the Coulomb case, which is critical as far as the computation of the energy is concerned. No shift arises for potentials decaying slightly faster than $1/r$ at infinity. 

In this paper we provide a simple and physically intuitive proof of the equality of the Jellium and UEG ground state energies in the thermodynamic limit, by explaining how to modify the floating crystal trial state. Our argument is to immerse the floating crystal into a thin layer of fluid. The small layer of fluid around the floating crystal is used to compensate the large charge fluctuations at the boundary of the system, which are responsible for the undesired shift of the energy.  The trial state suggests that the UEG ground state in infinite volume is indeed the uniform average of the Jellium crystal, as was believed. In a finite system, the particles of the UEG are probably not crystallized in a neighborhood of the boundary, however. 

Our argument will use a third definition of Jellium, which has always been of high relevance in practical computations~\cite{BruSahTel-66,Hansen-73,PolHan-73,CepChe-77,JonCep-96}. In this third point of view, the electrons are placed on a large torus without any background, whereas the Coulomb potential is replaced by a periodized version without zero mode. The problem of showing that the periodic system has the same thermodynamic limit has a long history for short range potentials~\cite{FisLeb-70,AngNen-73}. For the Coulomb potential, a rigorous proof seems to have been provided only recently, in a series of works~\cite{SanSer-15,PetSer-17,RouSer-16,LebSer-17,CotPet-19b}. For completeness we will also give a simple argument for this important fact. 

During the preparation of this work, the equality of the Jellium and UEG energies was claimed in a preprint~\cite[version~5]{CotPet-19b}. But the argument is long and indirect. Contrary to our trial state approach, it does not seem to provide any insight on the possible form of the UEG ground state. 

The paper is organized as follows. In the next section we introduce the three definitions. In Section~\ref{sec:crystal} we explain how to modify the floating crystal argument to prove an upper bound on the UEG energy. If Jellium was rigorously proved to be crystallized, this would complete the proof of the equality of Jellium and the UEG. In Section~\ref{sec:upper_UEG} we apply the modified floating crystal argument in the case of a unit cell of large but fixed side length $L$, containing $N=L^3$ electrons. After passing to the thermodynamic and taking $L\to\ii$ in a second step, this gives an upper bound on the UEG energy in terms of the periodic energy. Finally, in Section~\ref{sec:upper_per} we give a simple proof that the periodic problem coincides with the Jellium problem in the thermodynamic limit, which concludes the proof of the equality of the three definitions.  Section~\ref{sec:other_dim} contains a discussion on how our argument can be generalized to other interaction potentials and other space dimensions. We particularly consider the case of Riesz interaction potentials $r^{-s}$. 

\section{Three definitions of the ground state energy}

We only discuss here the classical case where the kinetic energy is dropped. We expect that a similar construction should apply to the quantum model but are unable to make this work at the moment. This is due to the Pauli principle which makes it difficult to merge two quantum systems with overlapping supports, as is explained in~\cite{LewLieSei-19} and is needed to add the thin layer of fluid around the Wigner crystal. 

By scaling we may assume in the classical case that the density is $\rho=1$. 
The Jellium energy of $N$ point charges in a background $\Omega_N\subset\R^3$ (a domain with volume $|\Omega_N|=N$) is given by
\begin{multline}
\cE_{\rm Jel}(\Omega_N,\bx_1,...,\bx_N)=\sum_{1\leq j<k\leq N}\frac{1}{|\bx_j-\bx_k|}\\
-\sum_{j=1}^N\int_{\Omega_N}\frac{\dy}{|\bx_j-\by|}
+\frac12 \iint_{\Omega_N\times\Omega_N}\frac{\dy\,\dz}{|\by-\bz|}.
\label{eq:Jellium_energy}
\end{multline}
For any given $\Omega_N$ we may minimize the energy over the positions $\bx_j$. It does not matter whether we constrain the point charges to stay in $\Omega_N$ or allow them to visit the whole space $\R^3$. After minimization they will always all end up in $\Omega_N$, since the energy is a harmonic function outside of $\Omega_N$ with respect to each $\bx_j$, when the other particles are fixed. It was also proved by one of us~\cite{Lieb-unp} that the point charges in $\Omega_N$ must have a universal positive distance to each other, a theorem that was recently used in~\cite{RouSer-16,LieRouYng-19}. We define the Jellium ground state energy per unit volume by
\begin{equation}
e_{\rm Jel}=\lim_{\Omega_N\nearrow\R^3}\min_{\bx_1,...,\bx_N\in\R^3}\frac{\cE_{\rm Jel}(\Omega_N,\bx_1,...,\bx_N)}{|\Omega_N|}.
\label{eq:Jellium_GS_energy}
\end{equation}
Under some natural technical conditions on $\partial\Omega_N$, the limit was proved to exist and to be independent of the sequence $\Omega_N$ in~\cite{LieNar-75}. The reader may think of $\Omega_N=N^{1/3}\Omega$ where $\Omega$ is a fixed open convex set of volume $|\Omega|=1$, for instance a cube or a ball. It is a famous conjecture~\cite{Wigner-34,Wigner-38} that the electrons crystallize on a BCC lattice, that is,
$$e_{\rm Jel}=\zeta_{\rm BCC}(1)\simeq -1.4442$$
where $\zeta_{\rm BCC}(s)$ is the Epstein Zeta function of the (density one) BCC lattice, see~\cite{ColMar-60,BlaLew-15} and~\cite[p.~43]{GiuVig-05}.

Next we turn to periodic Jellium, which is formally obtained when we repeat periodically a Jellium configuration in the whole space and compute its energy per unit volume. For simplicity we work with a cube (that is, we place the particles on the torus), but the argument is the same for other tilings. For $N=L^3$ we introduce
\begin{equation}
\cE_{{\rm per},L}(\bx_1,...,\bx_N)=\sum_{1\leq j<k\leq N}G_L(\bx_j-\bx_k)+\frac{N}{2L}M
\label{eq:Jellium_energy_periodic}
\end{equation}
where 
$$G_L(\bx)=\frac{G_1(\bx/L)}L=\frac{4\pi }{L^3}\sum_{\substack{\bk\in(2\pi/L)\Z^3\\ \bk\neq0}}\frac{e^{i\bk\cdot\bx}}{k^2}$$ 
with $G_1$ the $\Z^3$--periodic Coulomb potential, that is, the unique solution of the equation $-\Delta G_1=4\pi(\sum_{\bz\in \Z^3}\delta_\bz-1)$ such that $\int_{C_1}G_1=0$, with $C_1=(-1/2,1/2)^3$ the unit cube. The constant $M$ appearing in~\eqref{eq:Jellium_energy_periodic} is the Madelung constant of the cubic lattice which may be defined by
$$M=\lim_{\br\to0}\left(G_1(\br)-\frac{1}{r}\right).$$
In another point of view, $M/2=\zeta_{\Z^3}(1)$ is the Jellium energy per unit volume of the cubic lattice, that is, the interaction of each particle with all its periodic images. Except for the unimportant constant $M/(2L)$ which disappears in the thermodynamic limit, one can obtain~\eqref{eq:Jellium_energy_periodic} from~\eqref{eq:Jellium_energy} by replacing $1/r$ by the periodic function $G_L(\br)$ whenever $\Omega_N$ is a box. This is because $\int_{C_L}G_L=0$ hence the two background terms disappear. We define the ground state energy by
\begin{equation}
e_{\rm per}=\lim_{L\to\ii}\min_{\bx_1,...,\bx_N\in C_L}\frac{\cE_{{\rm per},L}(\bx_1,...,\bx_N)}{L^3}
\label{eq:Jellium_GS_energy_periodic}
\end{equation}
with $C_L=(-L/2,L/2)^3$. The limit on the right clearly exists when $L=2^nL_0$ because we can use as trial state a $2^nL_0$--periodic configuration in a cube of size $2^{n+1}L_0$, hence the right side is decreasing. The existence of the limit for $L\to\ii$ was proved in~\cite{SanSer-15,PetSer-17,RouSer-16,LebSer-17} but it will also be a consequence of our analysis. 

We finally turn to the UEG ground state energy. In this case there is no background but the electrons are assumed to form a constant charge density, say over a given set $\Omega_N\subset\R^3$. The \emph{indirect energy} of a given density $\rho$ with $\int_{\R^3}\rho(\br)\dr=N$ reads 
\begin{multline}
\cE_{{\rm Ind}}(\rho):=\min_{\rho_\bP=\rho}\int_{\R^{3N}}\sum_{1\leq j<k\leq N}\frac{{\rm d}\bP(\bx_1,...,\bx_N)}{|\bx_j-\bx_k|}\\
-\frac12 \iint_{\R^3\times\R^3}\frac{\rho(\bx)\,\rho(\by)}{|\bx-\by|}\dx\,\dy
\label{eq:UEG_energy}
\end{multline}
where the first minimum is taken over all $N$-particle probability measures $\bP$ with one-particle density $\rho$. Since the electrons are indistinguishable we should restrict ourselves to symmetric $\bP$'s, but the minimum is the same. Note that $\cE_{\rm Ind}(\rho)$ can be obtained from the Levy-Lieb functional of DFT~\cite{Levy-79,Lieb-83b} by taking $\hbar\to0$ or, equivalently, scaling the density in the manner $\lambda^3\rho(\lambda\bx)$ with $\lambda\to0$~\cite{Seidl-99,CotFriKlu-13,BinPas-17,Lewin-18,CotFriKlu-18,GroKooGieSeiCohMorGor-17}.
The ground state energy per unit volume of the UEG is given by
\begin{equation}
e_{\rm UEG}=\lim_{\Omega_N\nearrow\R^3}\frac{\cE_{\rm Ind}(\1_{\Omega_N})}{|\Omega_N|}. 
 \label{eq:UEG_GS_energy}
\end{equation}
It was proved in~\cite{LewLieSei-18} that the limit exists under the same conditions on $\Omega_N$ as for~\eqref{eq:Jellium_GS_energy}. 

One can replace the characteristic function $\1_{\Omega_N}$ by any sequence of densities $\rho_N$ which are equal to 1 well inside $\Omega_N$ (at a distance $\ell\ll|\Omega_N|^{1/3}$ from the boundary), equal to 0 well outside, and which stay bounded in the transition region. While such a $\rho_N$ is not exactly constant, we proved in~\cite{LewLieSei-18} that $\cE_{\rm Ind}(\rho_N)$ has the same thermodynamic limit as in~\eqref{eq:UEG_GS_energy}. We shall take advantage of this relaxed formulation in the following.

As shown in~\cite{LewLieSei-18,LewLieSei-19}, the constant $e_{\rm UEG}$ naturally arises in the Local Density Approximation of DFT. For instance, we have for a very spread out density in the form $\rho(\bx/N^{1/3})$ 
$$\lim_{N\to\ii}\frac{\cE_{\rm Ind}\big(\rho(\cdot/N^{1/3})\big)}{N}=e_{\rm UEG}\int_{\R^3}\rho(\bx)^{\frac43}\,\dx.$$
The classical UEG has been the object of many recent numerical works, based on methods from optimal transportation~\cite{Seidl-99,SeiPerLev-99,SeiGorSav-07,GorSei-10,SeiMarGerNenGieGor-17}. In addition to providing interesting properties of DFT at low density, the classical UEG has been used to get numerical bounds on the best constant in the Lieb-Oxford inequality~\cite{Lieb-79,LieOxf-80,LieSei-09,OdaCap-07,RasPitCapPro-09,LewLie-15}.

For any $N$-particle probability measure $\bP$ such that $\rho_\bP=\1_{\Omega_N}$, we have
\begin{align}
&\int_{\R^{3N}}\sum_{1\leq j<k\leq N}\frac{{\rm d}\bP(\bx_1,...,\bx_N)}{|\bx_j-\bx_k|}-\frac12 \iint_{\Omega_N\times\Omega_N}\frac{\dx\,\dy}{|\bx-\by|}\nn\\
&\qquad\qquad= \int_{\R^{3N}}\cE_{\rm Jel}(\Omega_N,\bx_1,...,\bx_N)\,{\rm d}\bP(\bx_1,...,\bx_N)\nn\\
&\qquad\qquad\geq \min_{\bx_1,...,\bx_N\in\R^3}\cE_{\rm Jel}(\Omega_N,\bx_1,...,\bx_N).\label{eq:compare_UEG_Jellium}
\end{align}
Hence, after optimizing over $\bP$ we obtain
\begin{equation}
\cE_{\rm Ind}(\1_{\Omega_N})\geq \min_{\bx_1,...,\bx_N\in\R^3}\cE_{\rm Jel}(\Omega_N,\bx_1,...,\bx_N).
\label{eq:easy_inequality_pre}
\end{equation}
After passing to the thermodynamic limit this yields the lower bound
\begin{equation}
e_{\rm UEG}\geq e_{\rm Jel}.
\label{eq:easy_inequality}
\end{equation}
The question of equality has been left open. Our main result is the following 
\begin{theorem}\label{thm:main}
We have $e_{\rm Jel}=e_{\rm per}=e_{\rm UEG}$.
\end{theorem}
The proof will be given in Sections~\ref{sec:upper_UEG} and~\ref{sec:upper_per}. 

\section{The floating crystal}\label{sec:crystal}
Before showing Theorem~\ref{thm:main} and as an illustration of the main idea,  we first prove that 
\begin{equation}
 e_{\rm UEG}\leq \zeta_{\rm BCC}(1)\simeq -1.4442.
 \label{eq:to_be_proved_crystal}
\end{equation}
If we had a proof that Jellium is crystallized in a BCC lattice, then this would immediately imply that $e_{\rm UEG}= e_{\rm Jel}$, due to~\eqref{eq:easy_inequality}. Note that~\eqref{eq:to_be_proved_crystal} also implies that the best constant in the Lieb-Oxford inequality~\cite{Lieb-79,LieOxf-80,LewLie-15} is at least as large as $-\zeta_{\rm BCC}(1)\simeq 1.4442$.

We first explain the floating crystal and the problem associated with its use as a trial state for estimating the UEG energy. 
We use the same notation as in~\cite[App.~B]{LewLie-15}. Let $\cL$ be the BCC lattice, with Wigner-Seitz unit cell $Q$ centered at 0, such that $|Q|=1$~\cite{AshcroftMermin}. We place the particles on the intersection of the lattice $\cL$ with a large cube $C'$ and call 
$\cL\cap C'=\{\bx_1,...,\bx_N\}$
the corresponding positions of the $N$ particles. We then take 
$$\Omega_N=\bigcup_{j=1}^N (Q+\bx_j),$$ 
the union of the cells centered at the particles. The floating crystal~\cite{BisLuh-82,MikZie-02,DruRadTraTowNee-04} is obtained by taking the delta function distribution of the $N$ particles, then translating by an amount $\ba\in\R^3$ and integrating $\ba$ over the unit cell $Q$. This corresponds to the $N$-particle probability 
\begin{equation}
 \tilde \bP=\int_Q \delta_{\bx_1+\ba}\otimes\cdots\otimes\delta_{\bx_N+\ba}\;\da
 \label{eq:floating_crystal}
\end{equation}
which has the constant density $\rho_{\tilde\bP}=\1_{\Omega_N}$.
The indirect energy per particle of this state is
\begin{equation}
\frac1{2N}\sum_{1\leq j<k\leq N}\frac{1}{|\bx_j-\bx_k|}-\frac1{2N}\iint_{\Omega_N\times\Omega_N}\frac{\dx\,\dy}{|\bx-\by|}.
\label{eq:indirect_background}
\end{equation}
In the limit $N\to\ii$, it has been shown in~\cite[App.~B]{LewLie-15} to converge to
\begin{equation}
\zeta_{\rm BCC}(1)+\frac{2\pi}{3}\int_Q x^2\,\dx\simeq-0.9507.
\label{eq:shift}
\end{equation}
By~\eqref{eq:compare_UEG_Jellium} the indirect energy per particle~\eqref{eq:indirect_background} can also be written in terms of a moving background in the form
\begin{equation}
\frac1N\int_{Q}\cE_{\rm Jel}(\Omega_N-\ba,\bx_1,...,\bx_N)\,\da.
\label{eq:average_Jellium}
\end{equation}
As explained in~\cite[App.~B]{LewLie-15}, the difference $\1_{\Omega_N-\ba}-\1_{\Omega_N}$ describes a monopole layer in a neighborhood of the surface which produces an electric potential felt by all the particles in the system. This survives in the thermodynamic limit and gives rise to the positive shift in~\eqref{eq:shift}. 

\begin{figure}[h]
\centering
\includegraphics[width=8cm]{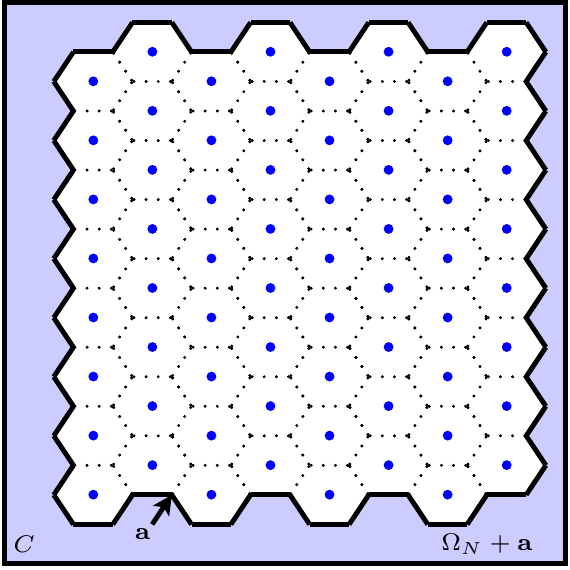}
\caption{A two-dimensional picture of the modified floating crystal~\eqref{eq:floating_crystal_bP} used in the text. The dots represent the point particles which are at the centers of hexagons of volume one. As the whole crystal block $\Omega_N$ is translated by $\ba$, the incompressible fluid gets displaced to fill the remaining space $C\setminus (\Omega_N+\ba)$. In other words, for any $\ba$, if the point charges were replaced by uniform charges over the hexagons, the total density would be equal to one over the whole box $C$. 
\label{fig:crystal}}
\end{figure}

We now explain our key new idea to avoid the energy shift. We immerse the crystal in a thin layer of fluid of density one. As we average over the positions of the crystal, the fluid gets displaced as depicted in Figure~\ref{fig:crystal}. 
To this end, we choose a slightly larger cubic container $C$ such that 
$\Omega_N+Q\subset C$
with the volume of the fluid $|C\setminus \Omega_N|=M$ being an integer. We will need $M\ll N$, so that the fluid layer around the floating crystal has a negligible volume in the thermodynamic limit. In practice, we choose the cube $C$ to be at a finite distance to the boundary of $\Omega_N$, this distance being larger than the diameter of $Q$. Then $M\sim N^{2/3}$. Our new trial state has the $N$ particles on the floating crystal, translated by $\ba\in Q$ as before, together with $M$ other particles forming a fluid in $C\setminus(\Omega_N+\ba)$, the set remaining after we have subtracted the \emph{moving} background:
\begin{equation}
 \bP=\int_Q \delta_{\bx_1+\ba}\otimes\cdots\otimes\delta_{\bx_N+\ba}\otimes \left(\frac{\1_{C\setminus (\Omega_N+\ba)}}{M}\right)^{\otimes M}\da.
 \label{eq:floating_crystal_bP}
\end{equation}
Note that the fluid is correlated with the position of the crystal. A sketch of the set-up is depicted in Figure~\ref{fig:crystal}. The density of this trial state equals
\begin{align}
\rho_{\bP}&=\int_Q\left(\sum_{j=1}^N \delta_{\bx_1+\ba}+\1_{C\setminus (\Omega_N+\ba)}\right)\,\da\nn\\
&=\1_{C}+\1_{\Omega_N}-\1_{\Omega_N}\ast\1_Q\label{eq:formula_rho}
\end{align}
with $f\ast g(\bx)=\int_{\R^3}f(\by)g(\bx-\by)\,\dy$ the convolution between two functions. In order to compute the Coulomb energy of $\bP$, it is convenient to denote the Hartree energy by
$$D(f,g):=\frac12 \iint_{\R^3\times\R^3}\frac{f(\bx)g(\by)}{|\bx-\by|}\dx\,\dy$$
and to use the shorthand notation $D(f):=D(f,f)$. 
Then we find
\begin{align*}
&\int_{(\R^3)^{N+M}}\sum_{1\leq j<k\leq N+M}\frac{1}{|\bz_j-\bz_k|}\,{\rm d}\bP(\bz_1,...,\bz_{N+M})\\
& = \sum_{1\leq i<j\leq N} \frac 1{|\bx_j - \bx_k|}+ \sum_{j=1}^N\int_Q  \int_{C\setminus (\Omega_N+\ba)} \frac{\da\,\dy}{|\bx_j +\ba-\by|}  \\
&\quad  +\left (1 - \frac 1{M}\right)  \int_Q \, D\left(\1_{C\setminus (\Omega_N+\ba)}\right)\,\da \\
& = \sum_{1\leq i<j\leq N} \frac 1{|\bx_j - \bx_k|}- \sum_{j=1}^N  \int_{\Omega_N} \frac{\dy}{|\bx_j  - \by|}+D(\1_{\Omega_N})\\
&\quad+  D(\1_C) + 2 D(\1_C,\1_{\Omega_N}-\1_{\Omega_N}\ast\1_Q)\\
& \quad     -  \frac 1{M} \int_Q \, D\left(\1_{C\setminus (\Omega_N+\ba)}\right)\,\da.
\end{align*}
The first line is the Jellium energy $\cE_{\rm Jel}(\Omega_N,\bx_1,...,\bx_N)$ of the finite crystal, whereas the second line is equal to $D(\rho_{\bP}) - D(\1_{\Omega_N}-\1_{\Omega_N}\ast\1_Q)$. Hence we have shown that the indirect energy of our trial state equals
\begin{align}
&\int_{(\R^3)^{N+M}}\sum_{1\leq j<k\leq N+M}\frac{{\rm d}\bP(\bz_1,...,\bz_{N+M})}{|\bz_j-\bz_k|}-D\big(\rho_{\bP}\big)\nn\\
&\qquad  = \cE_{\rm Jel}(\Omega_N,\bx_1,...,\bx_N) - D\big(\1_{\Omega_N}- \1_{\Omega_N}\ast \1_Q\big)\nn\\
& \qquad\qquad      -  \frac 1{M} \int_Q \, D\left(\1_{C\setminus (\Omega_N+\ba)}\right)\,\da.\label{eq:formula_floating_crystal_layer}
\end{align}
Using that $D(f)\geq0$, the last two terms can be neglected for an upper bound. We have therefore proved that 
\begin{equation}
 \cE_{\rm Ind}(\rho_{\bP})\leq \cE_{\rm Jel}(\Omega_N,\bx_1,...,\bx_N).
 \label{eq:upper_bd_UEG_lattice}
\end{equation}
The function $\rho_{\bP}$ is equal to 1 when $\bx$ is inside $\Omega_N$, at a distance at least equal to the diameter of $Q$ from the boundary $\partial\Omega_N$, whereas it is equal to 0 outside of $C$. It varies between 0 and 2 in the intermediate region. Since $M\ll N$ we can make use of the relaxed formulation in~\cite{LewLieSei-18} mentioned after~\eqref{eq:UEG_GS_energy} to conclude that 
$$\lim_{N\to\ii}\frac{\cE_{\rm Ind}(\rho_{\bP})}{N}=e_{\rm UEG}.$$
Therefore, after passing to the limit in~\eqref{eq:upper_bd_UEG_lattice}, we find the claimed upper bound
\begin{equation}
e_{\rm UEG}\leq \zeta_{\rm BCC}(1)\simeq -1.4442.
\label{eq:upper_bd_UEG_lattice_final}
\end{equation}

\section{Upper bound on the Uniform Electron Gas energy}\label{sec:upper_UEG}
The previous upper bound~\eqref{eq:upper_bd_UEG_lattice_final} is not enough to conclude that $e_{\rm UEG}=e_{\rm Jel}$ since we have no rigorous proof that Jellium is crystallized. However, the previous section contains the main idea. Here we show that 
\begin{equation}
e_{\rm UEG}\leq \frac{\cE_{\rm per,\ell}(\bx_1,...,\bx_{n}) }{n}
 \label{eq:upper_bd_UEG_periodic_positions}
\end{equation}
for any fixed $\bx_1,...,\bx_n$ points in the cube $C_\ell$, with $\ell^3=n$. The bound~\eqref{eq:upper_bd_UEG_lattice_final} simply corresponds to $n=1$ for the BCC lattice, but for simplicity we work here with the cubic lattice. After minimizing over the $\bx_j$ and passing to the thermodynamic limit $\ell\to\ii$, we obtain the inequality
\begin{equation}
 e_{\rm UEG}\leq e_{\rm per}.
 \label{eq:upper_bd_UEG_periodic}
\end{equation}
It has been shown in~\cite{CotPet-19b} based on results from~\cite{SanSer-15,PetSer-17,RouSer-16,LebSer-17} that $e_{\rm per}=e_{\rm Jel}$, hence this concludes the proof of the theorem. Since the proof in these works is quite long and delicate, we provide a simpler argument based on~\cite{LieNar-75} in the next section, for completeness. But here we concentrate on proving~\eqref{eq:upper_bd_UEG_periodic_positions} and~\eqref{eq:upper_bd_UEG_periodic}.

Let us consider $n$ distinct points $\bx_1,...,\bx_n$ inside the cube $C_\ell$ and denote $\tau=n^{-1}\sum_{j=1}^n \bx_j$ their center of mass.
If we shift the background by $\tau$, we obtain a configuration with no dipole moment:
$$\int_{\R^3}\by\left(\sum_{j=1}^n\delta_{\bx_j}(\by)-\1_{C_\ell+\tau}(\by)\right)\,\dy=0.$$
Next we repeat our configuration $\bx_j$ periodically in space and add a layer of fluid as before, in a background shifted by $\tau$. This is done as follows. We define the large cube of side length $\ell (2K+1)$
$$\Omega_N=\bigcup_{\substack{\bk\in\Z^3\\ |k_1|,|k_2|,|k_3|\leq K}}(C_\ell+\ell\bk)$$
with $N=\ell ^3(2K+1)^3$ and pick $C$ to be a slightly larger cube so that 
$\Omega_N+2C_\ell\subset C$
and $|C\setminus \Omega_N|=M\ll N$. Our trial state is, similarly as in the previous section, given by
$$\bP=\frac1{\ell^3}\int_{C_\ell}\!\! \bigotimes_{\substack{j=1,...,n\\ \bk\in\Z^3\\ |k_1|,|k_2|,|k_3|\leq K}}\!\!\!\!\delta_{\bx_j+\ell\bk+\ba}\otimes \left(\frac{\1_{C\setminus (\Omega_N+\ba+\tau)}}{M}\right)^{\otimes M}\!\!\da$$
and it has the density
$$\rho_{\bP}=\1_{C}+\frac1n \sum_{j=1}^n\1_{\Omega_N+\bx_j}-\1_{\Omega_N+\tau}\ast\frac{\1_{C_\ell}}{\ell^3}.$$
This density is again equal to one well inside $\Omega_N$ and 0 outside of $C$. Defining $\bx_{n+1},...,\bx_N$ to be the $\bx_j+\ell\bk$ with $\bk\neq0$ (ordered in any chosen fashion), the exact same calculations as in the previous section give
\begin{align}
&\int_{(\R^3)^{N+M}}\sum_{1\leq j<k\leq N+M}\frac{1}{|\bz_j-\bz_k|}\,{\rm d}\bP(\bz_1,...,\bz_{N+M})\nn\\
& = \sum_{1\leq i<j\leq N} \frac 1{|\bx_j - \bx_k|}\nn\\
&\qquad + \ell^{-3}\sum_{j=1}^N\int_{C_\ell}  \int_{C\setminus (\Omega_N+\tau+\ba)} \frac{\da\,\dy}{|\bx_j +\ba-\by|}  \nn\\
&\qquad  +\ell^{-3}\left (1 - \frac 1{M}\right)  \int_{C_\ell} \, D\left(\1_{C\setminus (\Omega_N+\tau+\ba)}\right)\,\da \nn\\
&=\cE_{\rm Jel}(\Omega_N+\tau,\bx_1,...,\bx_N)+D(\rho_{\bP})\nn\\
&\qquad -D\left (\frac{1}{n}\sum_{j=1}^n\1_{\Omega_N+\bx_j}-\1_{\Omega_N+\tau}\ast\frac{\1_{C_\ell}}{\ell^3}\right)\nn\\ 
& \qquad     -  \frac 1{M\ell^3} \int_{C_\ell} \, D\left(\1_{C\setminus (\Omega_N+\tau+\ba)}\right)\,\da.\label{eq:formula_periodic_layer}
\end{align}
Hence we obtain
$$\frac{\cE_{\rm Ind}(\rho_{\bP})}{N}\leq \frac{\cE_{\rm Jel}(\Omega_N+\tau,\bx_1,...,\bx_N)}{N}.$$
As before, when $N\to\ii$ and $M/N\to0$, the left side converges to $e_{\rm UEG}$. Since the repeated configuration has no dipole moment, it is a well-known fact that 
\begin{equation}
\lim_{N\to\ii}\frac{\cE_{\rm Jel}(\Omega_N+\tau,\bx_1,...,\bx_N)}{N}=\frac{\cE_{\rm per,\ell}(\bx_1,...,\bx_n)}{n}.
\label{eq:limit_towards_periodic}
\end{equation}
This concludes the proof of~\eqref{eq:upper_bd_UEG_periodic_positions}, hence of~\eqref{eq:upper_bd_UEG_periodic}.

For completeness, we briefly explain how to derive the limit~\eqref{eq:limit_towards_periodic}. We start with the upper bound, which turns out to be sufficient for our purpose. Since the points $\bx_1,\ldots,\bx_n$ are strictly inside $C_\ell$, the periodically repeated points are located at a positive distance from each other, independent of $N$. Then we replace the point charges by small uniform balls of radius $\eta$. By Newton's theorem, this does not change the interaction between the point charges, whereas the interaction with the background is increased. With $\chi_\eta(\br)=\eta^{-3}\1(4\pi r^3/3\leq\eta^3)$ we obtain 
\begin{multline*}
\cE_{\rm Jel}(\Omega_N+\tau,\bx_1,...,\bx_N)\\
\leq D\left(\sum_{j=1}^N\chi_\eta(\cdot-\bx_j)-\1_{\Omega_N+\tau}\right)-\frac{N}{\eta}D(\chi_1). 
\end{multline*}
The density in the parenthesis equals
$$\sum_{j=1}^N\chi_\eta(\cdot-\bx_j)-\1_{\Omega_N+\tau}=\sum_{\substack{\bk\in\Z^3\\ |k_1|,|k_2|,|k_3|\leq K}}f(\bx+\ell\bk)$$
with $f=\sum_{j=1}^n\chi_\eta(\cdot-\bx_j)-\1_{C_\ell+\tau}$. Passing to Fourier variables we can write
\begin{multline*}
D\left(\sum_{j=1}^N\chi_\eta(\cdot-\bx_j)-\1_{\Omega_N+\tau}\right)\\
=2\pi\int_{\R^3}\frac{|\widehat{f}(\bp)|^2}{p^2}\bigg|\sum_{\substack{\bk\in\Z^3\\ |k_1|,|k_2|,|k_3|\leq K}}e^{i \ell\bp\cdot\bk}\bigg|^2\dpp.
\end{multline*}
Note that $\int_{C_\ell}f(\bx)\,\dx=0$ and that $f$ has no dipole moment:
$$\int_{C_\ell}\br\,f(\br)\,\dr
=\frac1n\sum_{j=1}^n\bx_j-\tau=0.$$
Hence the function $|\widehat{f}(\bp)|^2/p^2$ is continuous and vanishes at $p=0$. On the other hand, the square of the Dirichlet kernel converges weakly to the Dirac comb of the dual lattice
\begin{align*}
&\frac1{(2K+1)^3}\bigg|\sum_{\substack{\bk\in\Z^3\\ |k_1|,|k_2|,|k_3|\leq K}}e^{i \ell p\cdot\bk}\bigg|^2\\
&\qquad=\frac1{(2K+1)^3}\prod_{\nu=1}^3\frac{\sin^2\big(\ell p_\nu(K+1/2)\big)}{\sin^2(\ell p_\nu/2)}\\
&\qquad\wto \left(\frac{2\pi}{\ell}\right)^3\sum_{\bp\in(2\pi/\ell)\Z^d}\delta_\bp
\end{align*}
as $K\to\ii$. Going back to configuration space, this gives the convergence 
\begin{multline*}
\lim_{N\to\ii}\frac1N D\left(\sum_{j=1}^N\chi_\eta(\cdot-\bx_j)-\1_{\Omega_N+\tau}\right)\\
=\frac1{2n} \int_{C_\ell}\int_{C_\ell}G_\ell(\bx-\by)f(\bx)\,f(\by)\,\dx\dy.
\end{multline*}
Passing then to the limit $\eta\to0$ using that 
\begin{align*}
&\frac12\iint_{\R^3\times\R^3}G_\ell(\bx-\by)\chi_\eta(x)\chi_\eta(y)\dx\,\dy\\
&\qquad=\frac12\iint_{\R^3\times\R^3}G_\ell(\eta(\bx-\by))\chi_1(x)\chi_1(y)\dx\,\dy\\ 
&\qquad =\frac{D(\chi_1)}{\eta}+\frac{M}{2\ell}+o(1),
\end{align*}
we obtain, as was claimed, the upper bound in~\eqref{eq:limit_towards_periodic}.

The proof of the lower bound in~\eqref{eq:limit_towards_periodic} is similar. It requires to estimate the error made in the interaction with the background when we replace the point particles by uniform balls. By Newton's theorem, there is no error when the ball is outside of the background. In case of an intersection the error can be bounded by 
$$2N\int_{r\leq \eta}\frac{\dr}{r}=4\pi N\eta^2,$$
a term which disappears in the thermodynamic limit since we take $\eta\to0$.

\section{Upper bound on the periodic energy}\label{sec:upper_per}

We give here a short proof of the inequality
\begin{equation}
 e_{\rm per}\leq e_{\rm Jel}.
 \label{eq:upper_bd_periodic_Jellium}
\end{equation}
Our strategy follows the one of \cite{LieNar-75}, which is based on the earlier work in \cite{LieLeb-72} and is also described in~\cite{LieSei-09}. In combination with~\eqref{eq:easy_inequality} and~\eqref{eq:upper_bd_UEG_periodic} this completes the proof of Theorem~\ref{thm:main}.

We are going to use the important fact that Newton's theorem holds in the periodic cell, for neutral systems. More precisely, if we have a radial charge distribution $\rho$ compactly supported in a ball $B_R$ and such that $\int_{B_R}\rho(\br)\dr=0$, then for $L$ large enough so that $B_R\subset C_L$ we claim that 
$\rho\ast G_L=\rho\ast r^{-1}$
in $C_L$. This is because $V=\rho\ast 1/r$ vanishes outside of $B_R$, by Newton's theorem, hence the periodized potential $\sum_{\bk\in\Z^d}V(\br+\ell\bk)$ solves the same equation as $\rho\ast G_L$ and we must have $\rho\ast G_L=\rho\ast1/r+K$ in $C_L$. The constant $K$ is found to vanish after integration over $C_L$. We also infer that $\rho\ast G_L$ vanishes on $C_L\setminus B_R$. 

We use the Swiss cheese theorem~\cite[Sec.~14.5]{LieSei-09} to cover the cube $C_L$ with many balls (all of integer volume) of sizes ranging from some fixed $\ell_0$ to the largest one of order $\ell$. The volume not covered by the balls is small compared to $L^3$ if $\ell$ is large, and in particular goes to zero relative to $L^3$ if $\ell \to \infty$ after $L\to\infty$ (for fixed $\ell_0$ or, more generally, if $\ell_0\ll \ell$). In each ball $B_n$, we place $N_n=|B_n|$ particles in the optimal Jellium configuration of the ball. The remaining $M=N-\sum_n|B_n|$ particles are placed uniformly in the left-over cheese $S=C_L \setminus \bigcup_n B_n$.
We obtain an upper bound on the minimal energy in the box
$$e_{\rm per}(C_L)=\min_{\bx_1,...,\bx_N\in C_L}\cE_{{\rm per},L}(\bx_1,...,\bx_N)$$
of the form
\begin{align*}
e_{\rm per}(C_L)\leq &\sum_{1\leq j<k\leq N-M}G_L(\bx_j-\bx_k) \\
&\quad +\sum_{j=1}^{N-M}\int_{S}G_L(\bx_j-\by)\dy\\
&\quad +\frac12\left(1-\frac{1}{M}\right)\iint_{S\times S}G_L(\bx-\by)\,\dx\,\dy
\end{align*}
where $\bx_1,...,\bx_{N-M}$ denote the positions of the $N-M$ particles in $\cup_n B_n$. We then use that $S=C_L\setminus\cup_n B_n$ and the fact that $\int_{C_L}G_L=0$. Discarding the term of order $1/M$ for an upper bound, we find that the right side is bounded above by
\begin{multline*}
\sum_{1\leq j<k\leq N-M}G_L(\bx_j-\bx_k) -\sum_n\sum_{j=1}^{N-M}\int_{B_n}G_L(\bx_j-\by)\dy\\
+\frac12\sum_{n,m}\iint_{B_n\times B_m}G_L(\bx-\by)\,\dx\,\dy 
\end{multline*}
This is exactly the energy obtained by putting point particles together with a uniform background over $\cup_n B_n$. Next we can average the particle configurations in each ball  $B_n$ over rotations. Due to Newton's theorem recalled above, this cancels the interactions between the systems in different balls. We obtain the upper bound
\begin{multline*}
e_{\rm per}(L)\leq \sum_n\bigg(\sum_{1\leq j<k\leq |B_n|}\widetilde{G_L}(\bx_j^{(n)}-\bx_k^{(n)}) \\-\sum_{j=1}^{|B_n|}\int_{B_n}\widetilde{G_L}(\bx_j^{(n)}-\by)\,\dy
+\frac12\iint_{B_n\times B_n}\widetilde{G_L}(\bx-\by)\,\dx\,\dy\bigg)
\end{multline*}
where $\bx_j^{(n)}$ denote the point charges in the ball $B_n$ and $\widetilde{G_L}$ denotes the average of $G_L$ over rotations of the ball $B_n$.
As an upper bound, we thus obtain the sum of the jellium energy in each ball $B_n$, with interaction kernel $\widetilde{G_L}(\bx-\by)$ in place of $|\bx-\by|^{-1}$. As $L\to \infty$, the former converges to the latter, however. Hence dividing by $L^3$ and taking the successive limits $L\to\infty$, $\ell\to \infty$ and $\ell_0\to \infty$, we arrive at the desired result. 

\section{Extension to Riesz potentials in all space dimensions}\label{sec:other_dim}
Our argument in Sections~\ref{sec:crystal} and~\ref{sec:upper_UEG} applies to more general potentials in any dimension, since we have essentially only used that the interaction has a positive Fourier transform, so that $D(f)\geq0$. Here we quickly describe how to generalize our findings to Riesz potentials, which are defined by
$$V_s(\br)=\begin{cases}
r^{-s}&\text{for $s>0$,}\\
-\ln r&\text{for $s=0$,}\\
-r^{-s}&\text{for $s<0$.}
\end{cases}$$
For instance, $s=1$ is the 3D Coulomb case which can also be considered in dimensions $d=1,2$. The case $s=0$ plays a central role in many situations. This is the natural interaction arising in random matrix theory for $d=1,2$~\cite{Forrester-10}. It also arises in the study of star polymer solutions, at least at short distances, see~\cite{WitPin-86} and~\cite[Sec.~5]{Likos-01}. It is very convenient to enclose all these important physical situations in the one-parameter family of Riesz interactions. This has been useful to better understand how the decay of correlations~\cite{AlaMar-85,Martin-88} and sum rules~\cite{GruLugMar-80,MarYal-80,GruLebMar-81,FonMar-84,Martin-88} depend on the decay of the potential, that is, the parameter $s$. It does not seem to be a well known fact that adding the background in the spirit of Wigner is a very robust method which, as we will demonstrate, works for all $-2\leq s<d$ in any dimension, and not only in the Coulomb case. 

We define the Jellium energy of $N$ point particles by
\begin{multline}
\cE_{{\rm Jel},d,s}(\Omega_N,\bx_1,...,\bx_N)=\sum_{1\leq j<k\leq N}V_s(\bx_j-\bx_k)\\
-\sum_{j=1}^N\int_{\Omega_N}V_s(\bx_j-\by)\,\dy
+\frac12 \iint_{\Omega_N\times\Omega_N}V_s(\bx-\by)\dx\,\dy
\label{eq:Jellium_energy_Riesz}
\end{multline}
and always assume $s<d$ to ensure the finiteness of the last two terms. Of course, no background is necessary in the short range case $s>d$. The following says that the system is thermodynamically stable for all $-2\leq s<d$. 

\begin{lemma}[Stability for Riesz potentials]\label{lem:stability_Jellium}
Let $d\geq1$ and $-2\leq s<d$. We have, for a universal constant $C(d,s)$,
\begin{multline}
 \cE_{{\rm Jel},d,s}(\Omega,\bx_1,...,\bx_N)\\ \geq -
 \begin{cases}
\dps C(d,s)N&\text{for $0<s<d$,}\\
\dps C(d,0)N&\text{for $s=0$ and $N=|\Omega|$,}\\
0&\text{for $-2\leq s<0$ and $N=|\Omega|$,}
 \end{cases}
 \label{eq:lower_bound}
\end{multline}
for every $\bx_1,...,\bx_N\in \R^d$ and every bounded open set $\Omega\subset\R^d$.
\end{lemma}

Many authors work under the constraint that $s\geq d-2$, but the Coulomb case $s=d-2$ is not a natural threshold in the family of Riesz potentials. Since the previous result does not seem to be well known, we provide a proof in Appendix~\ref{app:proof_stability}. A similar lower bound was previously derived in~\cite[App.~B.2]{CotPet-19b} for $0<s<d$.

We can now define the lowest Jellium energy in a given background $\Omega_N$ with $|\Omega_N|=N$ by
\begin{equation}
e_{{\rm Jel},d,s}(\Omega_N):=\min_{\bx_1,...,\bx_N\in\R^d}\frac{\cE_{{\rm Jel},d,s}(\Omega_N,\bx_1,...,\bx_N)}{|\Omega_N|}.
\end{equation}
This function is uniformly bounded from below, due to Lemma~\ref{lem:stability_Jellium}. We claim that it is also bounded from above for ``reasonable'' sets. To prove this we have to construct one trial state with an energy of order $N$. Taking the uniform average for all the points in $\Omega_N$, we find
\begin{equation*}
e_{{\rm Jel},d,s}(\Omega_N)\leq -\frac{1}{2N^2}\iint_{\Omega_N\times\Omega_N}V_s(\bx-\by)\dx\,\dy. 
\end{equation*}
For $s>0$ the right side is negative, proving that $e_{{\rm Jel},d,s}(\Omega_N)\leq0$. For $s<0$ the right side diverges to $+\ii$ as $N\to\ii$, and the uniform average is not a good trial state. Let us instead consider $\Omega_N$ to be the union of $N$ smaller cubes of size one ($\Omega_N$ can be made a cube if $N=K^3$ with $K\in\N$). In each of the small cube we put exactly one particle, which we average uniformly over its small cube only. This cancels exactly the background and we are just left with the self-energies of the small cubes:
\begin{equation}
e_{{\rm Jel},d,s}(\Omega_N)
\leq -\frac{1}{2}\iint_{C_1\times C_1}V_s(\bx-\by)\dx\,\dy. 
\label{eq:local_uniform}
\end{equation}
This is of order one as claimed. This argument applies to all $s<d$ and any $\Omega_N$ which can be partitioned into $N$ sets of volume one and uniformly bounded diameter. This leads us to conjecture that the Jellium model with Riesz interaction has a thermodynamic limit for all $-2\leq s<d$ in any dimension.

We would like to consider the corresponding energy
$$e_{\rm Jel}(d,s):=\lim_{\Omega_N\nearrow\R^d}e_{{\rm Jel},d,s}(\Omega_N)$$
where $\Omega_N$ is any reasonable sequence of domains like cubes or balls, with $|\Omega_N|\in\N$. The existence of this limit has been obtained for $s=d-2$ in any dimension in~\cite{Kunz-74,LieNar-75,SarMer-76} and for $d-2<s<d$ (resp.~$0\leq s<d$ for $d=1,2$) in~\cite{PetSer-17,LebSer-17,CotPet-19b}. To our knowledge no proof has yet been given for smaller values of $s$. In those cases we define $e_{\rm Jel}(d,s)$ by a $\liminf$ instead of a limit.

We then consider the indirect energy 
\begin{multline}
\cE_{{\rm Ind},d,s}(\rho):=\min_{\rho_\bP=\rho}\int_{\R^{dN}}\!\!\sum_{1\leq j<k\leq N}\!V_s(\bx_j-\bx_k){\rm d}\bP(\bx_1,...,\bx_N)\\
-\frac12 \iint_{\R^d\times\R^d}\rho(\bx)\,\rho(\by)\,V_s(\bx-\by)\dx\,\dy
\label{eq:UEG_energy_Riesz}
\end{multline}
which satisfies as in~\eqref{eq:easy_inequality_pre}
\begin{equation}
 \frac{\cE_{{\rm Ind},d,s}(\1_{\Omega_N})}{|\Omega_N|}\geq e_{{\rm Jel},d,s}(\Omega_N)
 \label{eq:compare_Riesz_easy}
\end{equation}
for every domain $\Omega_N$. Our upper bound~\eqref{eq:local_uniform} applies to the UEG as well, showing that $|\Omega_N|^{-1}\cE_{{\rm Ind},d,s}(\1_{\Omega_N})$ is uniformly bounded for ``reasonable'' sets. By following the proof of~\cite[Thm.~2.6]{LewLieSei-18}, based on the subadditivity of the indirect energy, one can show that the limit
$$e_{\rm UEG}(d,s):=\lim_{\Omega_N\nearrow\R^d} \frac{\cE_{{\rm Ind},d,s}(\1_{\Omega_N})}{|\Omega_N|}$$
exists and does not depend on $\Omega_N$, for regular-enough sequences $\Omega_N\nearrow\R^d$ and for all $-2\leq s<d$. 

Finally, we consider the periodic problem 
\begin{multline*}
e_{{\rm per},d,s}(C_L)=\min_{\bx_1,...,\bx_N\in C_L}\sum_{1\leq j<k\leq N}G_{d,s,L}(\bx_j-\bx_k)\\+\frac{M_{d,s}N}{2L^s} 
\end{multline*}
where $N=L^d$ and the periodic function $G_{d,s,L}$ has its Fourier coefficients equal to $c_{d,s}k^{s-d}$ for $0\neq \bk\in (2\pi/L)\Z^d$ and the appropriate constant $c_{d,s}$. Here $M_{d,s}=2{\rm sgn}(s)\zeta_{\Z^d}(s)$ is the corresponding Madelung constant. At $s=0$ the term $M_{d,s}/(2L^s)$ is replaced by $\zeta_{\Z^d}'(0)-\zeta_{\Z^d}(0)\ln L$.
The limit 
\begin{equation}
 e_{\rm per}(d,s)=\lim_{L\to\ii}\frac{e_{{\rm per},d,s}(C_L)}{L^d}
 \label{eq:def_periodic_Riesz}
\end{equation}
exists for subsequences in the form $L=2^nL_0$, due to the monotonicity of the energy. In the absence of a proof for general sequences we define $e_{\rm per}(d,s)$ by a liminf instead of a limit. 

The Coulomb case $s=-1$ in dimension $d=1$ is completely understood. It is well known that Jellium is crystallized~\cite{Baxter-63,Choquard-75,Kunz-74} and a calculation furnishes
$$e_{\rm Jel}(1,-1)=-\zeta(-1)=\frac1{12}.$$
On the other hand, it has been proved in~\cite{ColPasMar-15} that the floating crystal, defined similarly as in~\eqref{eq:floating_crystal}, is the \emph{exact} ground state of the indirect energy $\cE_{\rm Ind}(\1_{[-N/2,N/2]})$. Hence the computations in~\cite{LewLie-15} imply that there is an energy shift:
$$e_{\rm UEG}(1,-1)=e_{\rm Jel}(1,-1)+\frac1{12}=\frac16.$$
Jellium and the UEG differ at $s=-1$ in 1D. Other values of $s$ are considered in~\cite{Leble-15,DiMarino-19}.

Next we discuss the adaptation of the argument in Section~\ref{sec:upper_UEG} to Riesz potentials. Our result is the following.

\begin{theorem}
In space dimension $d\geq1$ we have
\begin{equation*}
e_{\rm Jel}(d,s)\leq e_{\rm UEG}(d,s)\leq e_{\rm per}(d,s)
\end{equation*}
for all $\max\left(0,d-4\right)<s<d$. There is equality for $0< s<d$ in dimensions $d=1,2$ and for $d-2\leq s<d$ in dimensions $d\geq3$.
\end{theorem}

The proof goes as follows. The first inequality is an immediate consequence of~\eqref{eq:compare_Riesz_easy}. For the second inequality we follow the argument in Section~\ref{sec:upper_UEG}. The computation~\eqref{eq:formula_periodic_layer} continues to hold for the potential $V_s$. The last two Hartree terms in this equation continue to be negative when $s>0$ because the Fourier transform of $V_s$ is positive. The first Hartree term on the second line of~\eqref{eq:formula_periodic_layer} is negative even for $-2\leq s\leq0$ because the function in the argument has a vanishing integral (see Appendix~\ref{app:proof_stability}). But the Hartree term on the third line is positive for $s<0$ and it has no particular sign for $s=0$. This term is a $O(M N^{-s/d})$, or $O(M\ln N)$ for $s=0$. In our case where $M\sim N^{1-1/d}$, the last term in~\eqref{eq:formula_periodic_layer} is a $o(N)$ under the condition that $s>-1$. 

The convergence to the periodic Jellium energy in~\eqref{eq:limit_towards_periodic} continues to hold under the condition that $d-s<4$ for configurations which have no dipole moment. The convergence of $\cE_{\rm Ind}(\rho_{\bP})/N$ to $e_{\rm UEG}(d,s)$ was established in~\cite{CotPet-19} for all $s>0$. Hence we obtain the inequality for $\max\left(0,d-4\right)<s<d$, as claimed. Should the last convergence hold for all $s\geq-2$, as we believe, then the same theorem would hold under the weaker assumption that $\max\left(-1,d-4\right)<s<d$. 

Finally, the equality $e_{\rm per}(d,s)=e_{\rm Jel}(d,s)$ is shown in~\cite{PetSer-17,LebSer-17,CotPet-19b} for $\max(0,d-2)< s<d$ (resp. $0\leq s<d$ in $d=1,2$).  Note that our proof that $e_{\rm per}(d,s)\leq e_{\rm Jel}(d,s)$ in Section~\ref{sec:upper_per} easily extends to $s=d-2$ in all dimensions, but not to other values of $s$, because it relies on Newton's theorem.

\section{Conclusion}
In this paper we have compared definitions of the minimum energy of the uniform electron gas and of jellium. For many years it has been an open problem to prove rigorously that they are the same to leading order in the volume since it was known that the obvious method for constructing the uniform gas definitely did not lead to the desired equivalence. We have succeeded in proving the equivalence, and thus provide a firm foundation for some aspects of density functional theory.

\appendix
\section{Proof of Lemma~\ref{lem:stability_Jellium} on the stability of Jellium with Riesz potentials}\label{app:proof_stability}
We start with $s>0$. Our idea is to replace $V_s$ by a smaller potential $0\leq V_{s,M}\leq V_s$ which is continuous at the origin and still has a positive Fourier transform. Then we use that, for this potential,
\begin{align}
&\sum_{1\leq j<k\leq N}V_{s,M}(\bx_j-\bx_k)-\sum_{j=1}^N\int_\Omega V_{s,M}(\bx_j-\by)\,\dy\nn\\
&\qquad\qquad+\frac{1}2\iint_{\Omega\times\Omega}V_{s,M}(\bx-\by)\,\dx\,\dy\nn\\
&\quad=\frac{1}2\iint_{\R^d\times\R^d}V_{s,M}(\bx-\by)\,{\rm d}\mu(\bx)\,{\rm d}\mu(\by)-\frac{N}{2}V_{s,M}(0)\nn\\
&\quad \geq -\frac{N}{2}V_{s,M}(0),
\label{eq:Onsager}
\end{align}
with $\mu=\sum_{j=1}^N\delta_{\bx_j}-\1_\Omega$. To define the potential $V_{s,M}$ we follow~\cite{FefLla-86,HaiSei-02} and first remark that for any radial function $\chi\geq0$ with $\int_{\R^d}\chi=1$,
\begin{equation}
\frac{1}{r^s}=c(s)\int_0^\ii \chi\ast\chi (t\br) \frac{{\rm d}t}{t^{1-s}}
\end{equation}
where
\begin{equation*}
c(s)^{-1}=\frac1{|S^{d-1}|}\iint_{\R^d\times\R^d}\frac{\chi(\bx)\chi(\by)}{|\bx-\by|^{d-s}} \dx\,\dy.
\end{equation*}
This suggests to introduce the truncated potential
$$V_{s,M}(\br):=c(s) \int_0^M \chi\ast\chi (t\br) \frac{{\rm d}t}{t^{1-s}}$$
which satisfies 
$V_{s,M}(0)=\frac{c(s)M^s}{s}\int_{\R^d}\chi^2$
and
$$\int_{\R^d}\big(V_s(\br)-V_{s,M}(\br)\big)\,\dr=\frac{c(s)M^{s-d}}{d-s}.$$
Using $V_{s,M}\leq V_{s}$ for the self energies of the particles and of the background, as well as
$$\sum_{j=1}^N\int_\Omega \big(V_s(\bx_j-\by)-V_{s,M}(\bx_j-\by)\big)\,\dy\leq N\int_{\R^d}\big(V_s-V_{s,M}\big)$$ 
for their mutual interaction, we find 
\begin{align*}
&\cE_{{\rm Jel},d,s}(\Omega,\bx_1,...,\bx_N)\\
&\geq \sum_{1\leq j<k\leq N}V_{s,M}(\bx_j-\bx_k)-\sum_{j=1}^N\int_\Omega V_{s,M}(\bx_j-\by)\,\dy\\
&\qquad+\frac{1}2\iint_{\Omega\times\Omega}V_{s,M}(\bx-\by)\,\dx\,\dy- N\frac{c(s)M^{s-d}}{d-s}\\
&=\frac12\iint_{\R^d\times\R^d}V_{s,M}(\bx-\by)\,{\rm d}\mu(\bx)\,{\rm d}\mu(\by)\\
&\qquad-Nc(s)\left(\frac{M^{s-d}}{d-s}+\frac{M^s}{2s}\int_{\R^d}\chi^2\right).
\end{align*}
Optimizing over $M$ we obtain
$$\cE_{{\rm Jel},d,s}(\Omega,\bx_1,...,\bx_N)\geq-\frac{Ndc(s)}{2s(d-s)}2^{\frac{s}{d}}\left(\int_{\R^d}\chi^2\right)^{1-\frac{s}{d}}.$$
Our conclusion is that
\begin{equation}
 \cE_{{\rm Jel},d,s}(\Omega,\bx_1,...,\bx_N)\geq -C_\chi(d,s)N
 \label{eq:lower_bound_Jellium_Riesz}
\end{equation}
where
\begin{equation*}
C_\chi(d,s)=\frac{d|S^{d-1}|}{2s(d-s)}2^{\frac{s}{d}}\frac{\norm{\chi}_{L^2}^{2-\frac{2s}d}\norm{\chi}_{L^1}^{\frac{2s}d}}{\iint_{\R^d\times\R^d}\frac{\chi(\bx)\chi(\by)}{|\bx-\by|^{d-s}} \dx\,\dy}. 
 \label{eq:def_constant_stability}
\end{equation*}
The best bound is obtained after optimizing over $\chi$ but here we keep it fixed for simplicity. 
In the limit $s\to0$, we have 
\begin{multline*}
\cE_{{\rm Jel},d,s}(\Omega,\bx_1,...,\bx_N)\\=\frac{(N-|\Omega|)^2-N}{2}+s\,\cE_{{\rm Je},d,0}(\Omega,\bx_1,...,\bx_N)+O(s^2)
\end{multline*}
whereas 
\begin{multline*}
C_\chi(d,s)= \frac1{2}+\frac{s}{2d}\bigg(1+\log 2\\
+\frac{\int_0^\ii (\chi\ast\chi)'(t)\log \left(t\int_{\R^d}\chi^2\right)\,{\rm d}t}{\int_{\R^d}\chi^2}\bigg)+O(s^2). 
\end{multline*}
When $N=|\Omega|$, the term $-N/2$ cancels on both sides of~\eqref{eq:lower_bound_Jellium_Riesz}. Hence we can divide by $s$ and pass to the limit $s\to0$. We get the claimed estimate for $s=0$, with the appropriate definition of $C_\chi(d,0)$. 

Finally, for $s<0$ we can write directly
$$\cE_{{\rm Jel},d,s}(\Omega,\bx_1,...,\bx_N)=-\frac1{2}\int_{\R^d}\int_{\R^d}|\bx-\by|^{|s|}{\rm d}\mu(\bx)\,{\rm d}\mu(\by).$$
The Fourier transform of $\mu$ has the behavior at the origin
$$\widehat{\mu}(\bk)=\frac{1}{(2\pi)^{d/2}}\Big(N-|\Omega|-i\bk\cdot \mathbf{P}\Big)+o(k)$$
where $\mathbf{P}=\sum_{j=1}^N\bx_j-\int_\Omega \bx\,\dx$ is the corresponding dipole moment. Under the assumption that $N=|\Omega|$ and $-2<s<0$, we obtain that $k^{-d-|s|}|\widehat{\mu}(\bk)|^2$ is integrable at the origin, and therefore we obtain 
$$\cE_{{\rm Jel},d,s}(\Omega,\bx_1,...,\bx_N)=c_{d,s}\int_{\R^d}\frac{|\widehat{\mu}(k)|^2}{k^{d+|s|}}\,dk$$
where 
$c_{d,s}=-(2\pi)^{\frac{d}2}2^{\frac{d}2-1-s}\Gamma\left(\frac{d-s}{2}\right)\Gamma\left(\frac{s}{2}\right)^{-1}$
is positive for $-2<s<0$ (but negative for $-4<s<-2$). 
At $s=-2$ we can compute directly
$\cE_{{\rm Jel},d,-2}(\Omega,\bx_1,...,\bx_N)=|P|^2$
and we conclude, as we have claimed, that $\cE_{{\rm Jel},d,s}(\Omega,\bx_1,...,\bx_N)\geq0$ for all $-2\leq s<0$ when $|\Omega|=N$. 

\bigskip

\noindent\textbf{Acknowledgments.} This project has received funding from the European Research Council (ERC) under the European Union's Horizon 2020 research and innovation programme (grant agreements AQUAMS No 694227 of R.S. and MDFT No 725528 of M.L.). Part of this work was done when the authors were visiting the Institut Henri Poincar\'e in Paris and the Erwin Schr\"odinger Institute in Vienna.


%

\end{document}